\input{aipcheck}

\documentclass[
    ,final            
  ]
  {aipproc}

\layoutstyle{6x9}

\begin{document}

\title{Black hole mass and variability in quasars}


\author{M. Wold}{
  address={Institute of Theoretical Astrophysics, University of Oslo, P.O. Box 1029 Blindern, N-0315 Oslo, Norway}
}

\author{M.S.  Brotherton}{
  address={Department of Physics and Astronomy, University of Wyoming, Laramie, WY 82072, USA}
}

\author{Z. Shang}{
  address={Department of Physics and Astronomy, University of Wyoming, Laramie, WY 82072, USA}
  ,altaddress={Department of Physics, Tianjin Normal University, 300074 Tianjin, China} 
}

\begin{abstract}
We report on a study that finds a positive correlation between black hole mass and variability amplitude in quasars.  Roughly 100 quasars at $z<0.75$ were selected by matching objects from the QUEST1 Variability Survey with broad-lined objects from the Sloan Digital Sky Survey. Black hole masses were estimated with the virial method using the broad H$\beta$ line, and variability was characterized from the QUEST1 light curves. The correlation between black hole mass and variability amplitude is significant at the 99\% level or better and does not appear to be caused by obvious selection effects inherent to flux-limited samples. It is most evident for rest frame time lags of the order a few months up to the QUEST1 maximum temporal resolution of about 2 years. The correlation between black hole mass and variability amplitude means that the more massive black holes have larger percentage flux variations. Over 2-3 orders of magnitude in black hole mass, the amplitude increases by $\sim0.2$ mag. A likely explanation for the correlation is that the more massive black holes are starving and produce larger flux variations because they do not have a steady inflow of gaseous fuel. Assuming that the variability arises from changes in the accretion rate Li \& Cao [8] show that flux variations similar to those observed are expected as a consequence of the more massive black holes having cooler accretion disks.
\end{abstract}

\maketitle


\section{Introduction}

Variability is characteristic for the quasar phenomenon and all quasars have continuum variations at some level. The variability is non-periodic and erratic in nature with typical changes of a few tenths of a magnitude in the rest frame UV-optical region. The variations (not including blazar-like variability due to relativistic beaming) occur on time scales from days up to decades and the variability amplitude increases with time scale (Hook et al.\ [2]; Cristiani et al.\ [3]; Collier \& Peterson [5]).   

Variability is an important diagnostic of the physical processes responsible for black hole activity. Variability is also the only method by which we can probe the smallest physical scales in AGN. Via resolution in the time domain we are able to resolve much smaller physical scales than what can be achieved with even the largest telescopes.  

The real causes of quasar variability  remain opaque, and in order to gain a better understanding of the quasar phenomenon it is useful to study how quasar variability depends on different physical parameters, such as black hole mass, accretion rate, luminosity, Eddington ratio etc. Here we correlate long-term quasar variability with black hole mass, details can be found in Wold, Brotherton \& Shang [11].

\section{Sample selection}

A sample of 104 quasars was formed by correlating broad-lined objects from the Sloan DR2 (Abazajian et al.\  [1]) with quasars from the QUEST1 Variability Survey (Rengstorf et al.\ [9]). The redshift range was restricted to $z<0.75$ in order to keep the broad H$\beta$ line within the SDSS spectral coverage. The restricted redshift range also contributes toward limiting time dilation effects in the sample as the time lag in the rest frame of the quasars, $\tau$, relates to the observed time lag as $\tau = \tau_{\rm obs} / (1+z)$. 

Quasar black hole masses were estimated from the Sloan spectra from the width of the H$\beta$ line and the continuum luminosity at 5100 {\AA} using the virial method scaling relationships described by Vestergaard \& Peterson [10].  Black hole mass and bolometric luminosity is plotted as a function of redshift in the left-hand panel of  Fig.~1. 

\begin{figure}
\includegraphics[height=.3\textheight]{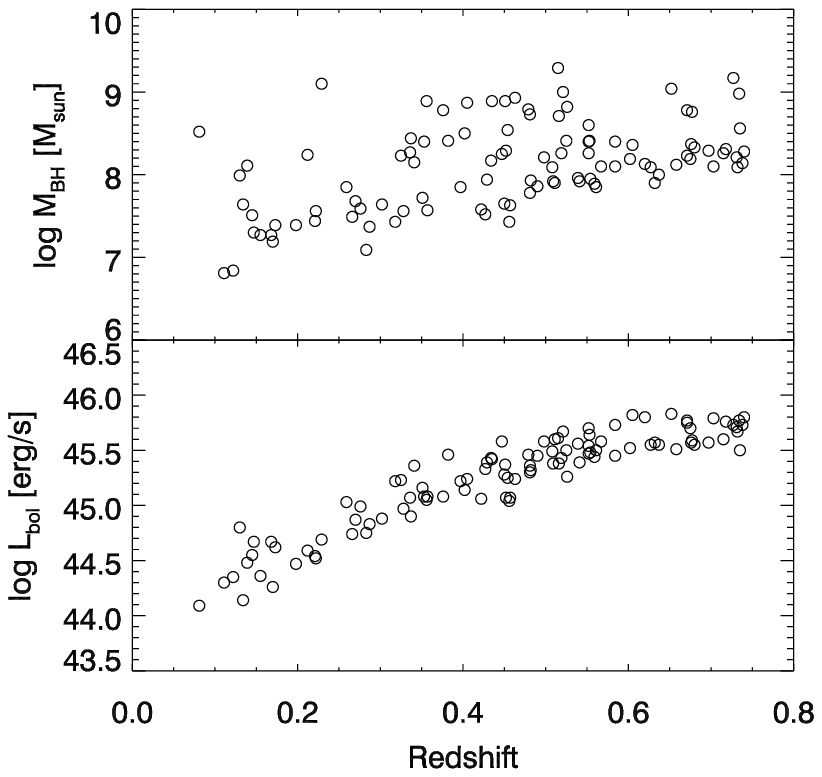}
\includegraphics[height=.3\textheight]{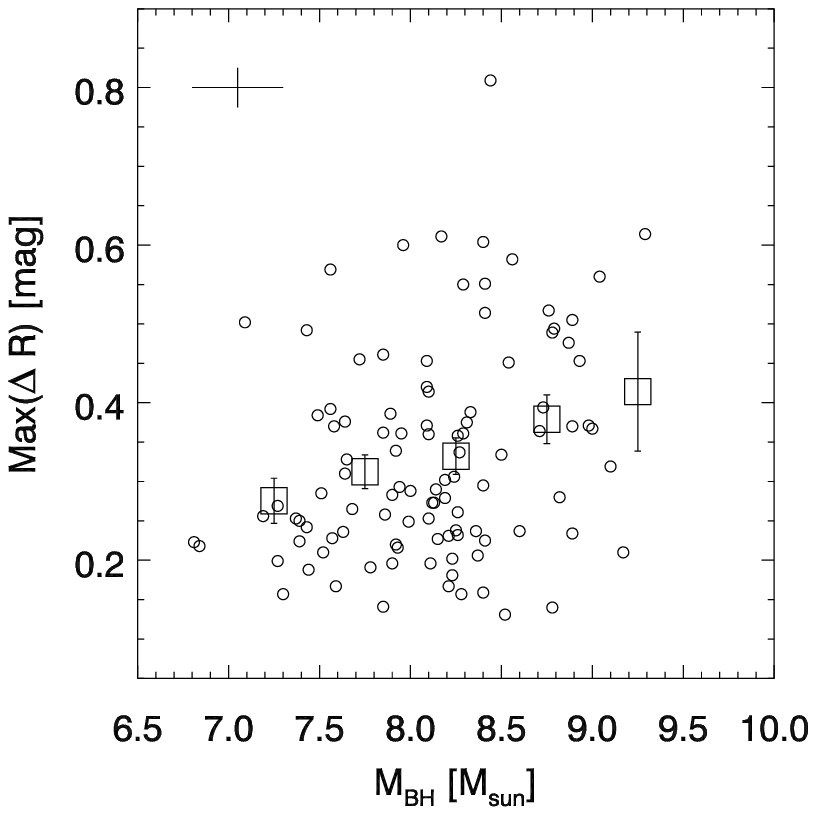}
\caption{{\bf Left:} Black hole mass and bolometric luminosity as a function of redshift for the SDSS-QUEST1 sample. {\bf Right:} Variability amplitude as a function of black hole mass. The open squares denote the mean in each black hole mass bin.}
\end{figure}

\section{Variability measurements}

We calculate the variability of each quasar from the distribution of all possible variability amplitudes (i.e. magnitude differences) on the quasar light curves, $\delta m_{ij}=m_{i}-m_{j}$, where $i<j$. The variability of one quasar is characterized by the standard deviation, mean, median and/or maximum of the distribution of $\delta m$ (see Giveon et al.\ [4]). The right-hand panel of Fig.~1 shows the variability defined in terms of the maximum magnitude difference on the light curves as a function of black hole mass. Over 2-3 orders of magnitude in black hole mass, the increase in variability amplitude is $\sim 0.2$ mag. 

We also calculate the variability for the whole sample as a function of time lag, referred to as the structure function $S(\tau) = <[m(t) - m(t+\tau)]^{2}>$, see left-hand panel of Fig.~2. The flat part of the structure function at time lags less than a few days is typical of the noise level in the data. The variations at longer time lags are caused by intrinsic quasar variations and are indeed seen to increase with time lag as expected. We isolate the (intrinsic) variations at longer time lags of $\tau > 100$ days and distribute them into bins of black hole mass. The result is shown in the right-hand panel of Fig.~2 where the increase in variability with black hole mass is seen. This figure also shows that it is the variability at time lags $>100$ days that contribute to the correlation.  

Peason correlation analysis shows that the correlation between black hole mass and variability is significant at the $\approx 3\sigma$ level. The strongest correlation is measured between black hole mass and maximum variability (correlation coefficient 0.285 and a two-sided probability of arising by chance of 0.3 \%). Other measures of variability have qualitatively similar, but somewhat less significant correlations with black hole mass. 

There are no significant correlations detected between variability and Eddington ratio, $L_{\rm bol}/L_{\rm Edd}$, possibly because of the small sample size. Wilhite et al.\ [6] using a larger sample of higher redshift Sloan quasars recently confirmed the black hole mass - variability correlation. Their data set is based on sampling a few points on the light curves of thousands of quasars, whereas our set is based on well-sampled light curves for a smaller number of quasars. 

 \section{Discussion}

We conclude that there is evidence for a real correlation between black hole mass and the long-term optical/UV variability amplitude of quasars. Our sample has redshift correlations typical of flux-limited samples as seen in the left-hand panel of Fig.~1. However, the black hole mass-variability correlation is not easily explained by such selection effects, discussed at length by Wold et al.\ [11]. We use partial correlation analysis to show that the black hole mass-variability correlation is present at constant luminosity, and also that the well-known anti-correlation between variability and luminosity exist in our sample.  

There are a number of physical arguments (see e.g.\ Collier \& Peterson [5] and references therein) that the characteristic variability time scale of quasars should depend on black hole mass. E.g. the characteristic variability time scale for accretion disk thermal changes is 1-2 years for black hole masses of $10^{8}$-$10^{9}$ M$_{\odot}$. Note however that we do not measure characteristic variability time scales. Rather we are measuring variability amplitudes on different time scales as a function of black hole mass. The ranges in black hole mass and characteristic variability time scale for thermal changes are well covered by our sample, hence we may be biased toward detecting variations caused by accretion disk thermal changes in quasars having black hole masses of $10^{8}$-$10^{9}$ M$_{\odot}$.  An explanation for the correlation could then be that we are preferentially detecting variability in quasars with more massive black holes. Quasars with lower mass black holes may have shorter characteristic time scales which escape the more well-sampled time scales in our study and thereby artificially create/enhance the correlation. 

A more likely explanation is that quasars with more massive black holes are starving, having swallowed already a large amount of gas and dust in their vicinity. The larger percentage flux variations for the more massive black holes could indicate that the black hole is running out of gaseous fuel. Because there is not a steady inflow of gaseous fuel, larger fluctuations in the continuum are seen. Quasars with more massive black holes may therefore experience larger changes in the accretion rate.  This would qualitatively explain our results. 

Furthermore, standard optically thick, geometrically thin accretion disks (Shakura \& Sunyaev [7]) are colder for larger black hole masses, and Li \& Cao [8] show that this also leads to an increase in variability amplitude with black hole mass, provided the variability is caused by changes in the accretion rate. For accretion rate changes of 40\% they show that amplitude changes corresponding to those we observe are expected over a range in black hole mass of approximately $10^{7}$-$10^{9}$ M$_{\odot}$. 

The black hole mass-variability correlation may therefore be caused by a combination of the latter two effects: 1) larger black holes have larger accretion rate changes and 2) larger black holes have cooler accretion disks. 


\begin{figure}
\includegraphics[height=.3\textheight]{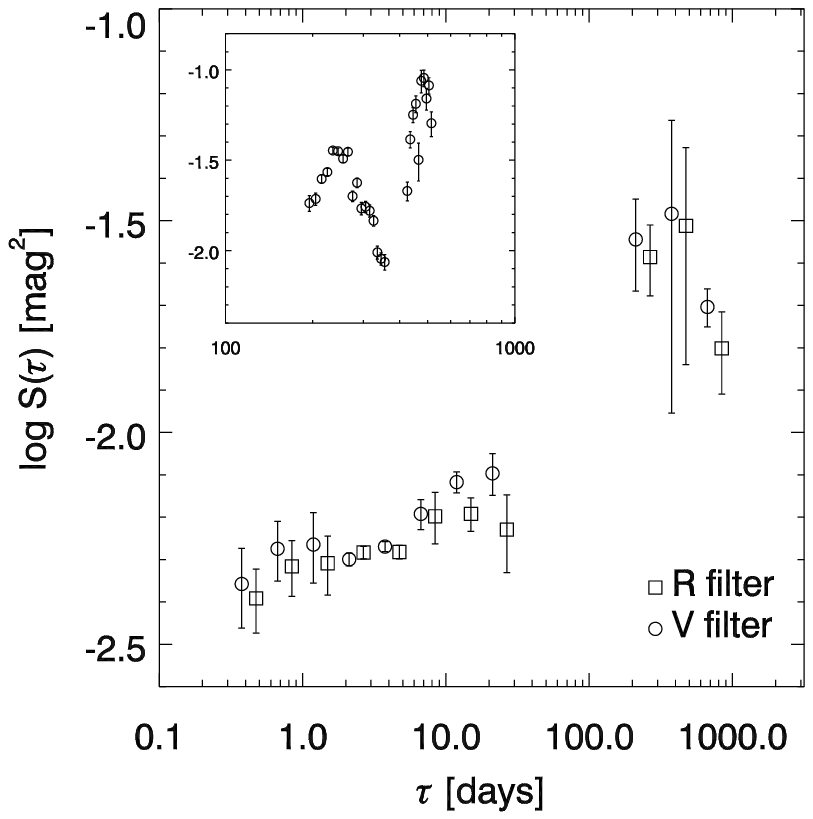}
\includegraphics[height=.3\textheight]{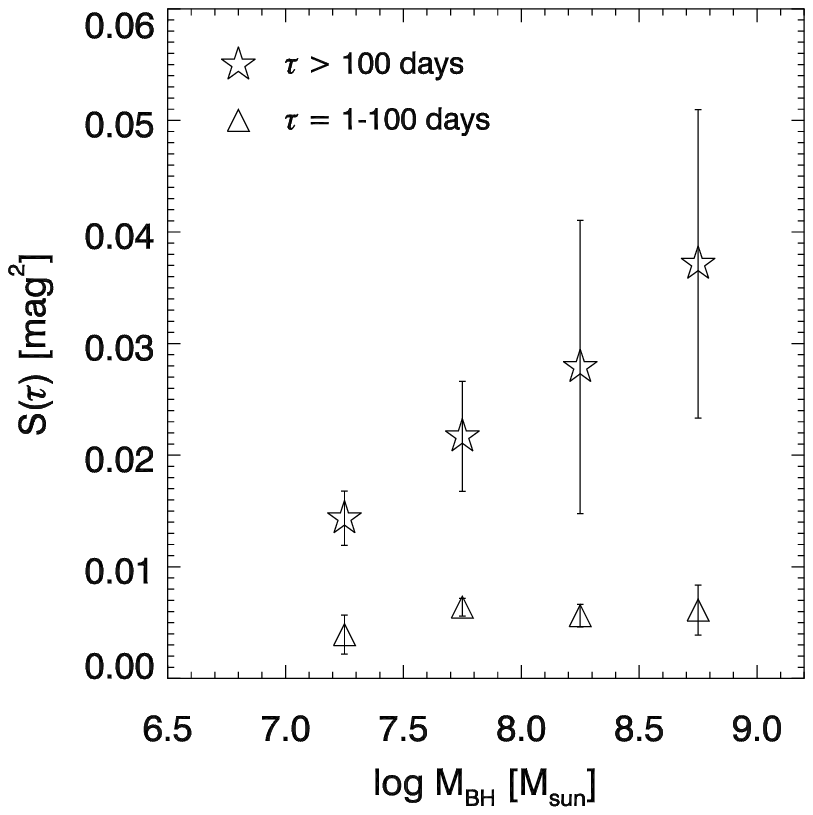}
\caption{{\bf Left:} Variability as a function of time lag (the structure function) for the whole sample. {\bf Right:} The structure function evaluated for four different bins in black hole mass and for two different time scales.}
\end{figure}





\bibliographystyle{aipproc}   




\end{document}